\documentclass[a4paper,10pt,twoside]{cpc-hepnp}
\usepackage[numbers,sort&compress]{natbib}
\usepackage{stmaryrd}
\usepackage{bbding}
\usepackage{amssymb}
\usepackage{multicol}
\usepackage{multirow}
\usepackage{graphicx}
\usepackage{subfigure}
\usepackage{booktabs}
\usepackage{amssymb,bm,mathrsfs,bbm,amscd}
\usepackage[tbtags]{amsmath}
\usepackage{lastpage}
\usepackage{array}
\usepackage{color}
\usepackage{epstopdf}

\begin{document}
%\begin{CJK*}{UTF8}{gbsn}

%\begin{CJK*}{GBK}{song}

\bibliographystyle{unsrt}

%\fancyhead[co]{\footnotesize XIU Qinglei et al.: Study of the Beamstrahlung Effects at the CEPC}
%\fancyhead[c]{\small Chinese Physics C~~~Vol. XX, No. X (201X)
%XXXXXX} \fancyfoot[C]{\small 010201-\thepage}

%\footnotetext[0]{Received XX-XX-XXXX}

\title{Study of the Beamstrahlung Effects at the CEPC
\thanks{The study was partially supported by the CAS/SAFEA International Partnership Program for Creative Research Teams, funding from CAS and IHEP for the Thousand Talent and Hundred Talent programs, as well as grants from the State Key Laboratory of Nuclear Electronics and Particle Detectors.}
}

\author{%
\quad Xiu Qinglei$^{1,2;1)}$\email{xiuql@ihep.ac.cn} %
\quad Zhu Hongbo$^{1,2;2)}$ \email{zhuhb@ihep.ac.cn}%
\quad Lou Xinchou$^{1,2,3}$
}
\maketitle

\address{%
$^1$ State Key Laboratory of Particle Detection and Electronics, Beijing 100049, China\\
$^2$ Institute of High Energy Physics, CAS, Beijing 100049, China\\
$^3$ University of Texas at Dallas, Richardson, TX 75080-3021, USA\\
}

\begin{abstract}
The discovery of a 125~GeV Higgs boson at the LHC marked a breakthrough in particle physics. The relative lightness of the new particle inspires the consideration of a high luminosity Circular Electron Positron Collider (CEPC) as a Higgs Factory to study the Higgs boson in a clean environment. At the CEPC, the beamstrahlung might represent one of the most important sources of beam-induced backgrounds that will impact the detector. It will introduce additional backgrounds to the CEPC detector through the subsequent electron-positron pair production and the hadronic process. Therefore its impacts should be carefully evaluated. In this paper, the beamstrahlung-induced backgrounds are first estimated with analytical methods and are further evaluated in detail with Monte Carlo simulation. The detector occupancy due to the beamstrahlung at the location where the first vertex detector layer may be placed is found to be well below 0.5\%. Radiation levels characterised as  non-ionising energy loss (NIEL) and total ionising dose (TID) are estimated to be $\sim 10^{11}~1~\mathrm{MeV}~n_{eq}/\mathrm{cm}^2/$yr and  $\sim$300 kRad/yr, respectively.

\end{abstract}

\begin{keyword}
CEPC, Beamstrahlung, Detector backgrounds, Pair production
\end{keyword}

\begin{pacs}
29.20.db, 29.27.-a, 25.20.Lj, 29.40.Gx
\end{pacs}

\begin{multicols}{2}

\section{Introduction}
After the Higgs boson was discovered by the ATLAS and CMS experiments~\cite{atlas:2012obs,cms:2012obs} at the CERN Large Hadron Collider (LHC), it becomes important to measure the Higgs boson properties with high precision, beyond the ultimate reach of the LHC~\cite{ATL-PHYS-PUB-2014-016}. This goal is generally believed to be only achievable at future lepton colliders (so-called ``Higgs Factory''). The proposed Circular Electron Position Collider (CEPC) will primarily operate at the center-of-mass energy range of 240 -- 250~GeV (near the $ZH$ production threshold) with an instantaneous luminosity of $2 \times 10^{34}~\mathrm{cm}^{-2}\mathrm{s}^{-1}$~\cite{cepc-sppc:2015pcdr}. To achieve the desired high luminosity, both electron and positron beams will have to be heavily squeezed to very small bunch sizes just before collision. At the interaction point (IP), trajectories of the electrons or positrons in one bunch are bent by the electromagnetic field of the beam particles of the opposite charge inside a crossing bunch. During this process, one special kind of synchrotron radiation, called ``beamstrahlung''\cite{Augustin:1978ah}, will be emitted. Subsequent processes including the electron-positron pair production and hadronic event generation~\cite{Schulte:1997tk}, both of which involve the energetic beamstrahlung photons, are considered among the most important detector backgrounds at the CEPC and should be carefully evaluated.

\section{Analytical Estimation of the Beamstrahlung Backgrounds}

\begin{table*}[htb]
\centering
\caption{Machine parameters of the LEP2, CEPC and ILC colliders and their average beamstrahlung parameters $\Upsilon_{av}$ calculated with Eq.~\ref{eq:avBSPar}.}.
\begin{tabular}{|p{5.5cm}|p{2.5cm}<{\centering}|p{1.5cm}<{\centering}|p{1.5cm}<{\centering}|p{1.5cm}<{\centering}|p{1.5cm}<{\centering}|}
\hline
\textbf{Parameters} & \textbf{Symbol} \ & \textbf{LEP2} & \textbf{CEPC} & \textbf{ILC250} & \textbf{ILC500} \\
\hline
\small{Center of mass energy} & $E_{cm}$~[GeV] & 209 & 240 & 250 & 500\\
%\hline
\small{Bunch population} & $N$~[$\times 10^{10}$] & 58 & 37.1 & 2 & 2\\
%\hline
\small{Horizontal beam size at IP} & $\sigma_x$~[nm] & 270000 & 73700 & 729 & 474\\
%\hline
\small{Vertical beam size at IP} & $\sigma_y$~[nm] & 3500 & 160 & 7.7 & 5.9\\
%\hline
\small{Bunch length} & $\sigma_z$~[$\mu$m] & 16000 & 2260 & 300 & 300\\
%\hline
\small{Horizontal beta function at IP} & $\beta_x$~[mm] & 1500 & 800 & 13 & 11\\
%\hline
\small{Vertical beta function at IP} & $\beta_y$~[mm] & 50 & 1.2 & 0.41 & 0.48\\
%\hline
\small{Normalized horizontal emittance at IP} & $\gamma\epsilon_x$~[mm $\cdot$ mrad] & 9.81 & 1594.5 & 10 & 10\\
%\hline
\small{Normalized vertical emittance at IP} & $\gamma\epsilon_y$~[mm $\cdot$ mrad] & 0.051 & 4.79 & 0.035 & 0.035\\
%\hline
\small{Luminosity} & $L$~[$10^{34}~ \mathrm{cm}^{-2} \mathrm{s}^{-1}$] & 0.013 & 1.8 & 0.75 & 1.8\\
\hline
\textbf{Beamstrahlung parameter} & $\bm{\Upsilon_{av}~ [\times 10^{-4}]}$ & $\bm{0.25}$ & $\bm{4.7}$ & $\bm{200}$ & $\bm{620}$\\
\hline

\end{tabular}

\label{tab:beamParameters}
\end{table*}

The beamstrahlung is usually characterised by the beamstrahlung parameter $\Upsilon$ :
\begin{equation}\label{eq:BSPar}
  \Upsilon = \frac{2}{3} \frac{h \omega_c}{E}
\end{equation}

\noindent where $\omega_c = \frac{3}{2} \gamma^{3} c/\rho$ denotes the critical energy of synchrotron radiation, $\rho$ the bending radius of the particle trajectory and $E$ the beam particle energy before radiation. The higher the $\Upsilon$, the more beamstrahlung photons with higher energies will be emitted. Assuming Gaussian charge distributions for the colliding beams, the average $\Upsilon$ can be estimated with the following formula:

\begin{equation}\label{eq:avBSPar}
  \Upsilon_{av} \approx \frac{5}{6} \frac{N r_{e}^{2} \gamma}{\alpha (\sigma_{x} + \sigma_{y}) \sigma_{z}}
\end{equation}

\noindent where $r_e$ is the classical electron radius, $\gamma$ the Lorentz factor of the beam particles, $\alpha$ the fine structure constant, $\sigma_{x}$/$\sigma_{y}$ the transverse size of the bunch and $\sigma_{z}$ the bunch length.

With the machine parameters of LEP2~\cite{Telnov:2012rm}, CEPC~\cite{cepc-sppc:2015pcdr} and ILC~\cite{Adolphsen:2013kya} listed in Table \ref{tab:beamParameters}, the average beamstrahlung parameters $\Upsilon_{av}$ of these colliders are calculated with Eq.~\ref{eq:avBSPar}. To achieve the required high luminosities, the bunch sizes of the CEPC and the ILC are expected to be much smaller than that of the LEP2, resulting in non-negligible beamstrahlung effects at the two machines.

\subsection{Beamstrahlung Photons}

Assuming Gaussian charge distributions for the two colliding bunches, the average number of photons  $n_\gamma$ emitted by each beam particle in one bunch crossing (BX) can be approximated as:
\begin{equation}\label{eq:nPhoton}
n_\gamma \approx 2.59 \left( \frac{\alpha \sigma_z}{\lambda_e \gamma} \Upsilon_{av} \right) \frac{1}{(1+\Upsilon_{av}^{2/3})^{1/2}}
\end{equation}

\noindent where $\lambda_e$ is the Compton wave length of the electron \cite{Yokoya:1991qz}. With the CEPC beam parameters, there will be only $\sim$0.22 photons emitted by an individual beam particle. However the total number of photons per bunch crossing can become significant because of the high bunch population of $3.71 \times 10^{11}$.

\subsection{Pair Production}
There are two kinds of processes for the pair creation in the beam-beam interactions, namely coherent and incoherent pair productions. In the coherent pair production, real photons convert into electron-positron pairs in the strong macroscopic electromagnetic field of the other crossing bunch. The total number of coherent pairs in each bunch crossing amounts to:
%$$n_b \approx \left( \frac{\alpha \sigma_z}{\gamma \lambda_e} \Upsilon_{max} \right)^{2}\Xi(\Upsilon_{max})$$
$$n_b \approx \left( \frac{\alpha \sigma_z}{\gamma \lambda_e} \Upsilon \right)^{2}\Xi(\Upsilon)$$
with
\begin{equation}\label{eq:nofCoherentPairs}
\Xi(\Upsilon) =
\begin{cases}
(7/128)\exp(-16/(3 \Upsilon))& (\Upsilon \lesssim 1)\\
0.295\Upsilon^{-2/3}(\ln\Upsilon - 2.488)& (\Upsilon \gg 1)
\end{cases}
\end{equation}

Eq.~\ref{eq:nofCoherentPairs} suggests that the coherent pair production with small $\Upsilon$ will be exponentially suppressed \cite{Chen:1992ax}. The number of coherent pairs produced per bunch crossing at the CEPC is very close to 0.

In the incoherent pair production, electron-positron pairs are produced by the interaction between two incoming photons, which could be real and/or virtual. There are three incoherent processes: the Breit-Wheeler process $\gamma\gamma \to e^{+} e^{-}$, in which both photons are real; the Bethe-Heitler process $e\gamma \to ee^{+} e^{-}$, in which one photon is real and the other is virtual; and the Landau-Lifshitz process $ee \to eee^{+} e^{-}$, in which both photons are virtual. The approximated cross sections of the three processes can be calculated with the formulae in Refs.~\cite{Yokoya:1991qz,Berestetskii:1971lp,Baier:1980kx}. There will be roughly 44 pairs from the Breit-Wheeler process, 327 pairs from the Bethe-Heitler process and 1322 pairs from the Landau-Lifshitz process produced in every bunch crossing at the CEPC.

\begin{figure*}[htb]
\begin{center}
\includegraphics[width=0.85\textwidth]{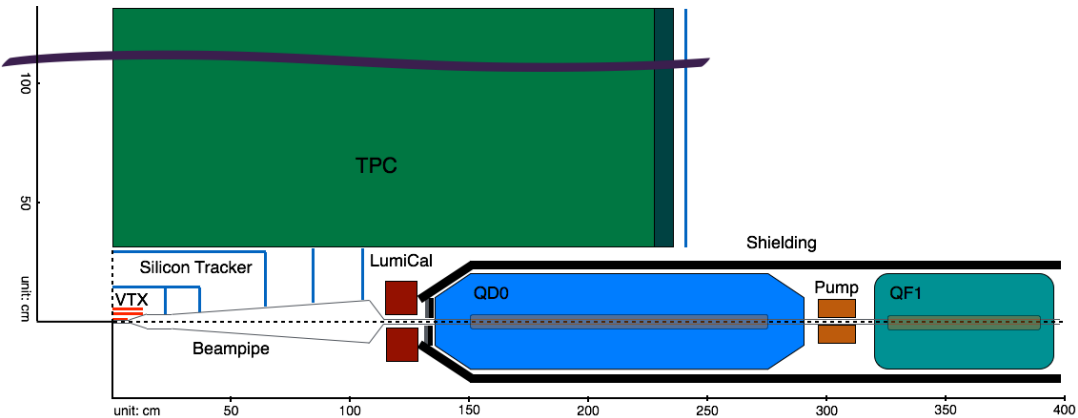}
\figcaption{\label{fig:CEPCLayout} {A possible layout of the CEPC interaction region.}}
\end{center}
\end{figure*}

\subsection{Hadronic Backgrounds}
In addition to the electron-positron pairs, two colliding photons can also produce hadrons with certain probability. This process is non-negligible because a small fraction of the events may contain final state particles of high transverse momenta. The cross section of the hadronic process can be parameterised as~\cite{Schuler:1996en}:

\begin{align}\label{eq:hadron}
\sigma_H & = \nonumber \\
& 211 \mathrm{nb}~ \cdot \left( \frac{s}{\mathrm{GeV}^2} \right)^{0.0808} + 297 \mathrm{nb}~ \cdot \left( \frac{s}{\mathrm{GeV}^2} \right)^{-0.4525}
\end{align}

\noindent where $s$ is the square of the center-of-mass energy of two photons. The number of hadronic events per bunch crossing at the CEPC will be much less than 1.

The analytical estimation suggests that there will be significant amount of beamstrahlung photons per bunch crossing at the CEPC. The consequent incoherent electron-positron pair production might be the dominant background induced by the beamstrahlung, while the coherent pair production and hadronic events could be negligible. Nevertheless, to fully assess their impacts on the CEPC detector, studies based on full Monte Carlo simulation should be performed.

\section{Simulation Studies of the Beamstrahlung Backgrounds}

The beam-beam interactions have been simulated with Guinea-Pig++ (Generator of Unwanted Interactions for Numerical Experiment Analysis Program Interfaced to GEANT)~\cite{Schulte:2007zz}, which allows detailed studies of the emission of the beamstrahlung photons, the incoherent pair production and the hadronic events. In the simulation, the charge distributions in both collision bunches are assumed to be Gaussian. The background events generated with Guinea-Pig++ are interfaced to Mokka~\cite{MoradeFreitas:2002kj}/GEANT4~\cite{Agostinelli:2002hh} for full detector simulation. Hit information such as hit density and total ionizing dose are extracted accordingly. %%using Marlin~\cite{Gaede:2006pj}, which is a framework for reconstruction and analysis.

The primary background particles hitting the beampipe or any detector component can cause particle shower. These secondary particles may enter the detector and become important background for the detector. Therefore the impact of beam backgrounds on the detector is highly depending on the layout and material budget of the beampipe and the detector. Fig.~\ref{fig:CEPCLayout} shows the interaction region layout used in this study, which features a short focal length of $L^{\ast}=1.5$~m, \textit{i.e.} the distance between the final quadrupole QD0 and the IP.

\section{Simulation Results}

\begin{center}
\includegraphics[width=8cm]{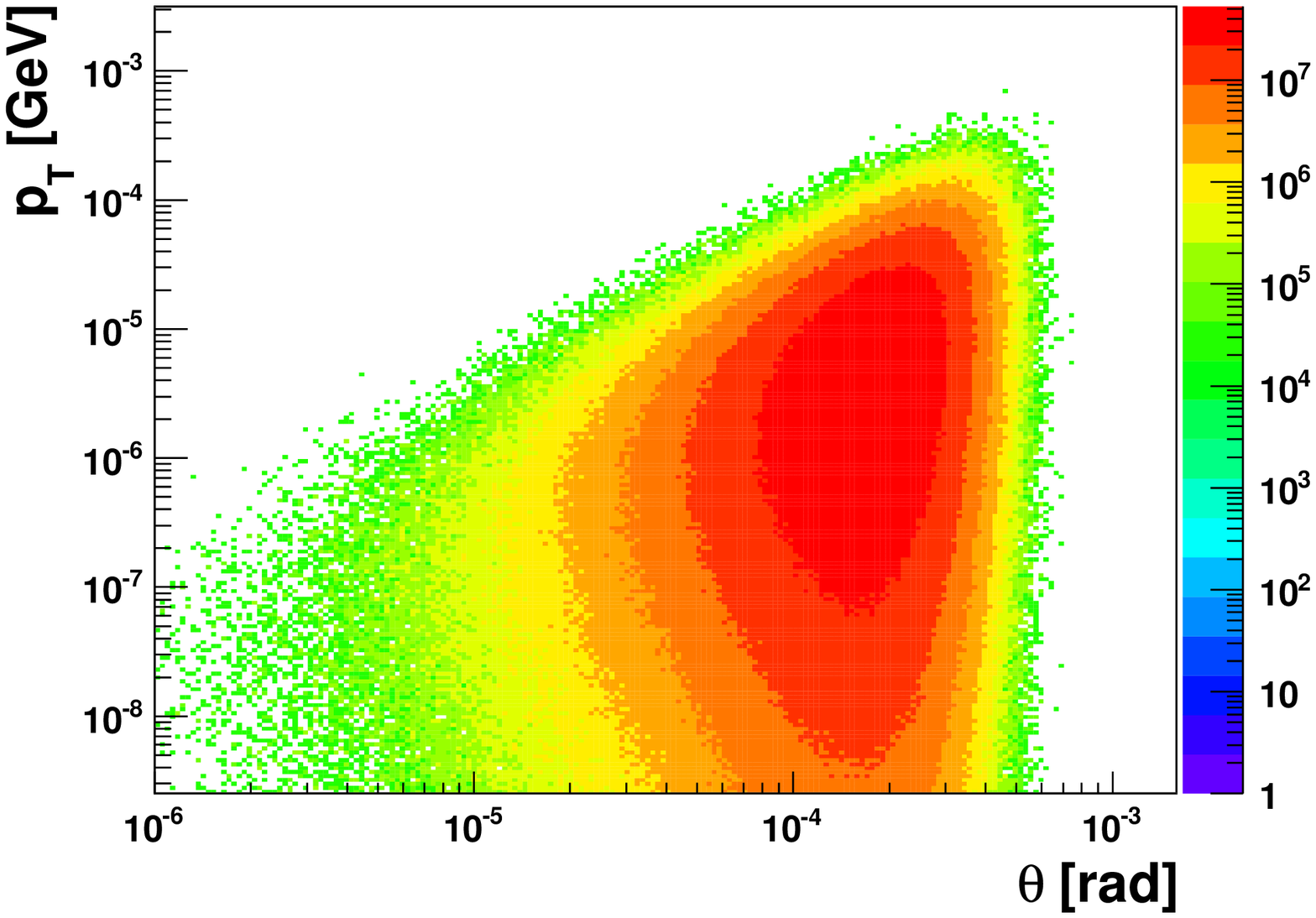}
\figcaption{\label{fig:PhotonPtThetaNormalizedCEPC} $p_{T}$ versus $\theta$ distribution of the beamstrahlung photons at the CEPC.}
\end{center}

\subsection{Beamstrahlung Photons}

Fig.~\ref{fig:PhotonPtThetaNormalizedCEPC} shows the transverse momentum distribution versus the polar angle of the beamstrahlung photons originating from the beam-beam interactions. The photons are confined within a small angle of $|\theta|<1$~mrad, which indicates that most of them will leave the interaction region without interacting with the beam pipe or any other detector component. Therefore they are not the major source of detector backgrounds.

\subsection{Incoherent Electron-Positron Pairs}

Unlike the beamstrahlung photons, the electrons and positrons from the incoherent pair production are usually produced with large polar angle and high transverse momentum. They can contribute to the detector backgrounds in direct and indirect ways:
%might increase the occupancy of the detector in two ways:
\begin{itemize}
	\item If the primary particles are with large enough polar angle and transverse momentum, they will hit the detector directly;
    \item If the primary particles are with small polar angle, they might still hit the beampipe or the downstream detector components and produce secondary particles, which might be back-scattered into the detector.
	%%The pairs with smaller polar angle will hit the beam pipe or down-stream detectors and produce lots of backscattered particles. Some of these backscattered particles may feed back to the inner part of the detector.
\end{itemize}

Fig.~\ref{fig:PairsPtThetaNormalizedCEPC} shows the $p_T$ versus $\theta$ distribution of the electrons and positrons. There is an empty region at the bottom-right corner of the distribution plot because electrons and positrons with energy below 5~MeV, which will be constrained in the beam pipe by the solenoid magnetic field, are not tracked in the simulation with Guinea-Pig++. It should be noted that the beam pipe and any detector component must be kept away from the particles populated in the red region to avoid particle showering. It requires additional optimisations to the beam pipe design and the downstream detector components layout in the interaction region.

\begin{center}
\includegraphics[width=8cm]{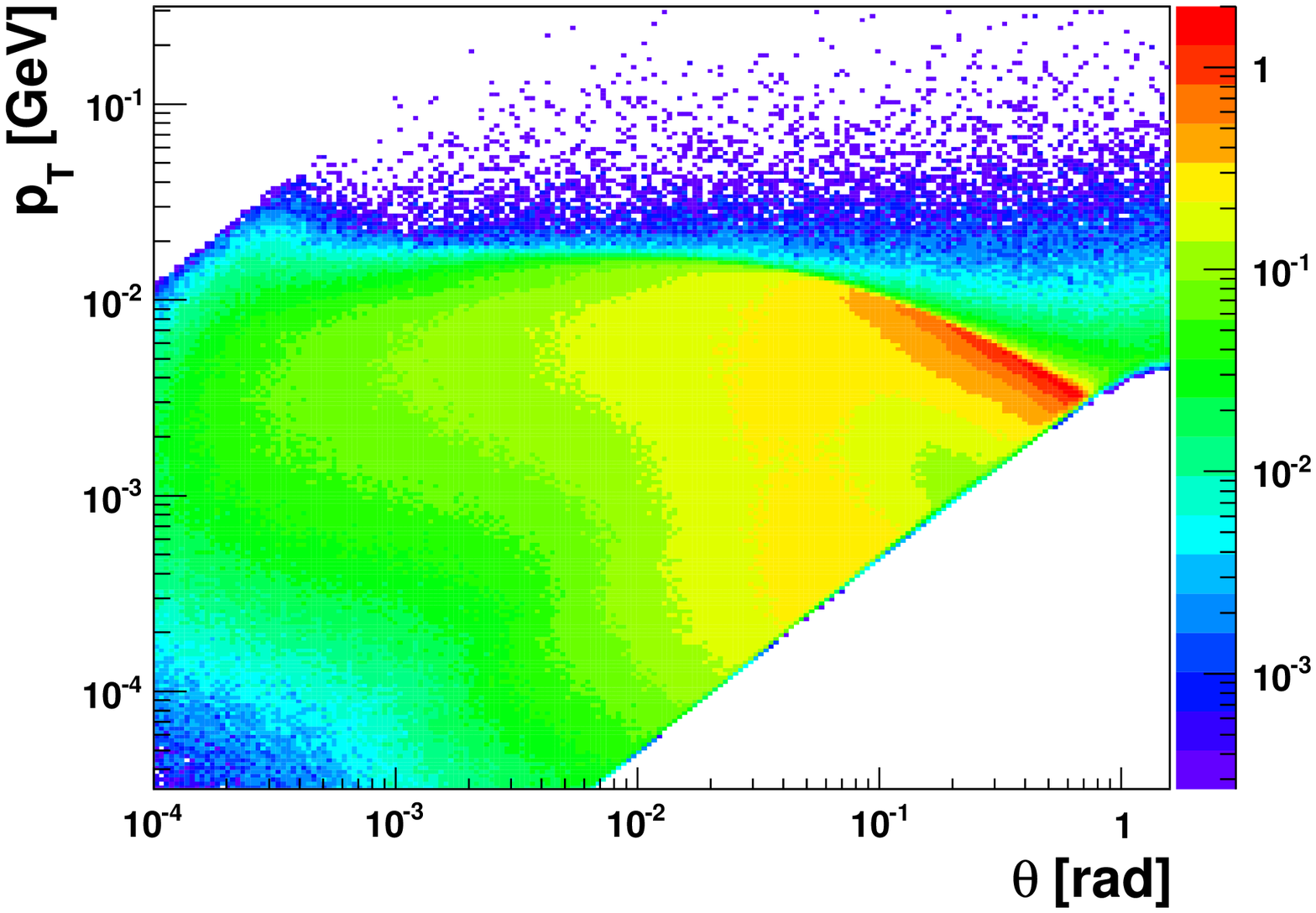}
\figcaption{\label{fig:PairsPtThetaNormalizedCEPC} The $p_{T}$ versus $\theta$ distribution of the electrons and positrons from the incoherent pair production.}
\end{center}

In the $p_{T}$ versus $p_{z}$ distribution of the electrons and positrons, the major fraction of the particles are concentrated in the area under the envelope (so-called ``kinematic edge''). The developed envelop can be fitted to an imperial formula as the following:

\begin{equation}\label{eq:Envelope}
p_T = 0.0202~ p_{Z}^{0.297}
\end{equation}

Assuming perfect helical trajectories in a solenoid magnetic filed of 3.5~T, the particles on the envelop can develop a trajectory profile as shown in Fig.~\ref{fig:PairHelixEdge}. The beam pipe, labeled as the red line, must be kept away from the helical trajectories.

\begin{center}
\includegraphics[width=8cm]{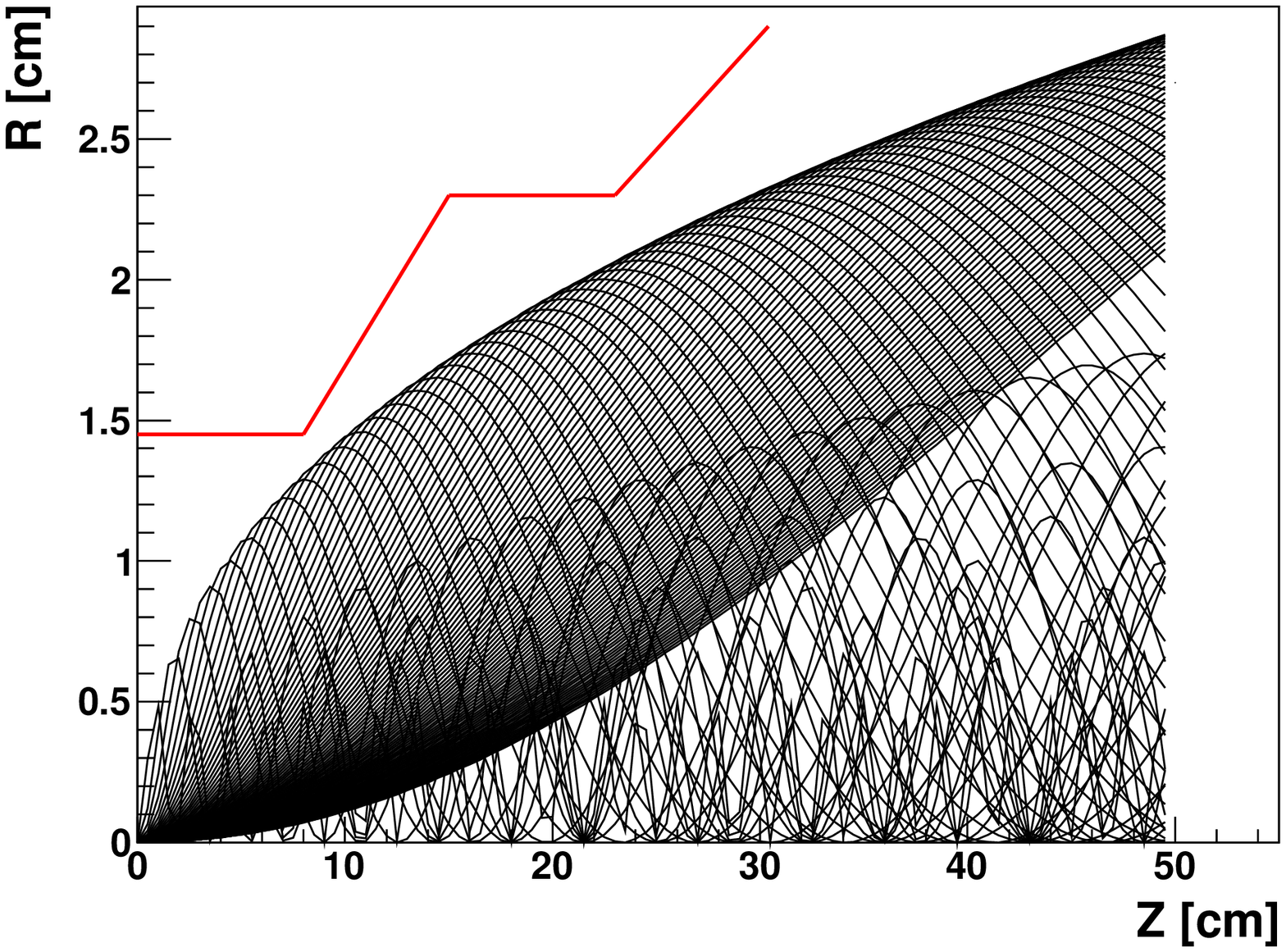}
\figcaption{\label{fig:PairHelixEdge} The helix trajectories of electrons/positrons from the pair production in a solenoid magnetic field of 3.5~T. The red line indicates the position of the beam pipe, which must be kept away from the profile of the helical trajectories.}
\end{center}

\subsection{Hit Density}
Although most of the particles originating from the beamstrahlung will leave the interaction region without hitting the beampipe or any detector component, a small fraction of them will enter the detector directly, or hit the beampipe and/or detector components in the forward region and introduce back-scattered particles. These primary and secondary particles will increase the detector occupancy and introduce radiation damage to the silicon detectors close to the interaction point.

\begin{center}
\includegraphics[width=8cm]{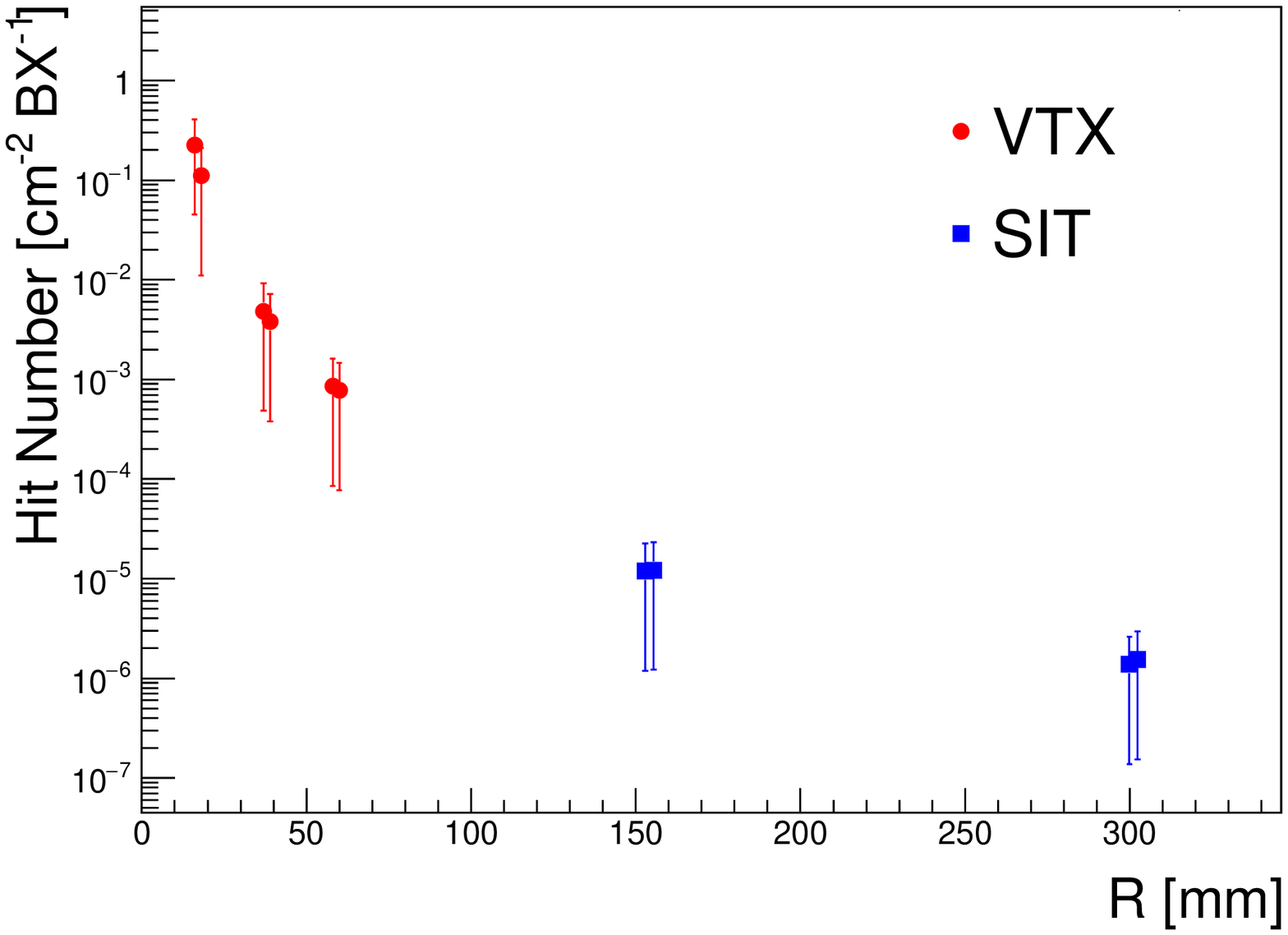}
\figcaption{\label{fig:HitDensityVXDSIT} Hit density at the VTX and SIT detector layers}
\end{center}

The estimated hit densities at the barrel layers of the vertex detector (VTX) and the silicon inner tracker (SIT) are shown in Fig.~\ref{fig:HitDensityVXDSIT}. On the inner most VTX layer at $r=1.6$~cm, the hit density amounts to 0.2~$\mathrm{hits} / \mathrm{cm}^2 /\mathrm{BX}$. With the design bunch spacing of $\sim3.6~ \mu$s for the CEPC, the corresponding detector occupancy will be well below 0.5\% for the first VTX layer with an assumed pixel pitch size of  20~$\mu$m and a readout time of 20~$\mu$s.

\subsection{Radiation Levels}
Although the estimated hit density is low per bunch crossing, the detector radiation levels can be still considerable given the high repetition frequency ($\sim 2.8 \times 10^{5}$ bunch crossings per second) of the CEPC machine. Their impacts on the CEPC detector, in particular the vertex detector close to the interaction point, should be carefully evaluated. The radiation damage in silicon detector can be roughly characterized as the non-ionizing energy loss(NIEL) and the total ionizing dose(TID).

NIEL can lead to crystal defects by displacing the silicon atom out of their lattice sites (so-called ``bulk damage'')~\cite{Newman:1982rp}. The effects induced by any particle with a given energy can be normalised to the equivalent damage caused by 1 MeV neutrons. In this study, the electron flux at a given point can be obtained by tracking all particles with perfect helixes. The NIEL per year is calculated with the same method used in Ref.~\cite{Fretwurst:2002vb}, assuming machine operation for $10^7$~seconds (Snowmass year) and with a safety factor of 10 taken into account. Fig.~\ref{fig:CEPCNIEL} shows the NIEL distribution in the CEPC vertex detector. At the first vertex detector layer, the annual value for NIEL is about $10^{11}~\mathrm{1~MeV}~ n_{eq}/\mathrm{cm}^2$/yr.

\begin{center}
\includegraphics[width=8cm]{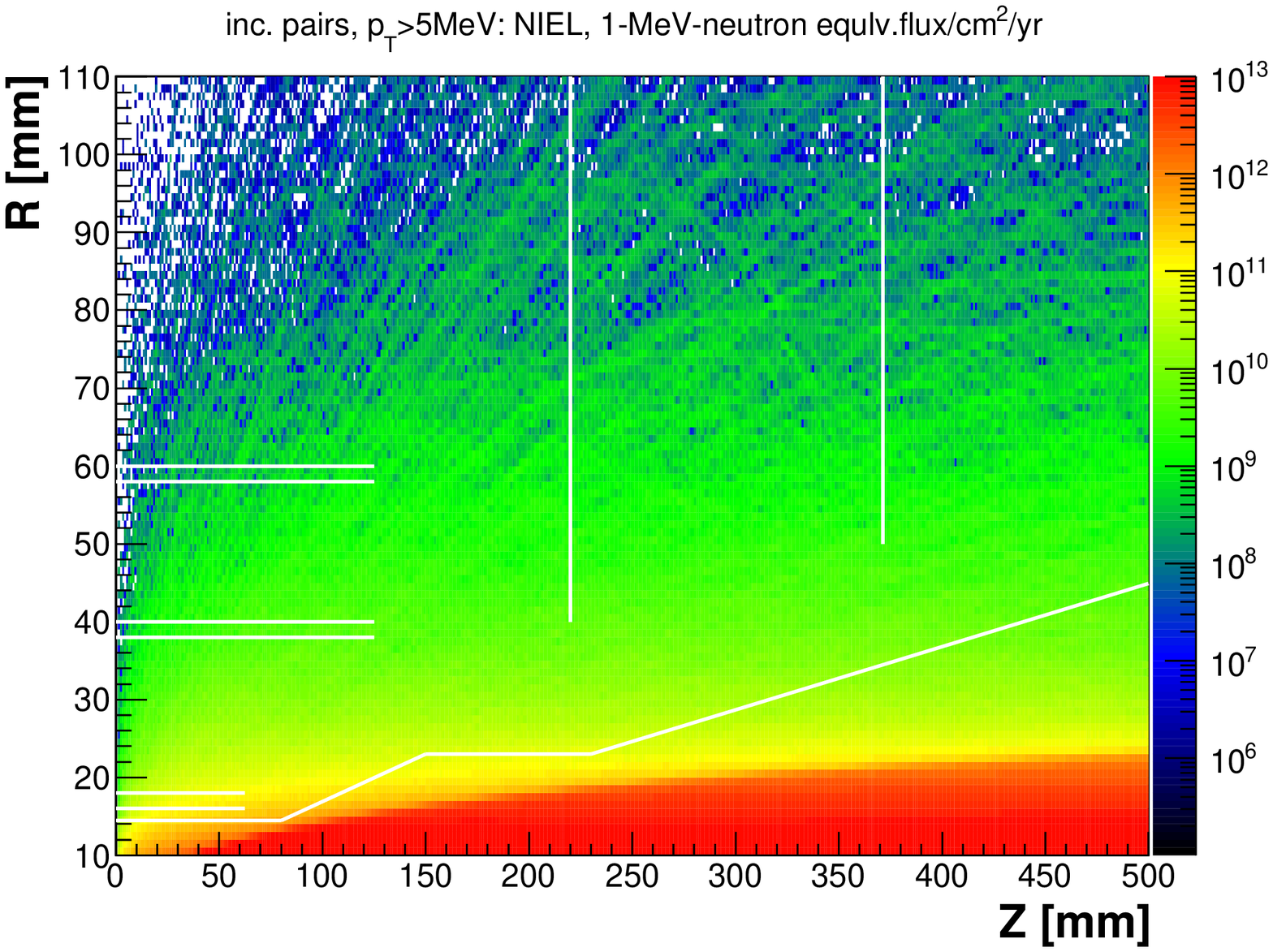}
\figcaption{\label{fig:CEPCNIEL} NIEL distribution in the CEPC vertex detector.}
\end{center}

On the other hand, TID can create ionization at the Si-SiO$_2$ interface and degrade the performance of silicon devices~\cite{Oldham:2003ns}. TID can be determined with:

\begin{equation}\label{eq:TID}
D = \frac{E_{dep}}{m} = \frac{E_{dep}}{\rho \cdot V}
\end{equation}

\noindent where $E_{dep}$ is the energy deposited in the matter, and $m$, $\rho$ and $V$ are the mass, the density and the volume of the matter, respectively. The annual TID in the VTX with full simulation is shown in Fig.~\ref{fig:CEPCVXDTID} and a safety factor of 10 is included in the calculation. The TID is estimated to be $\sim$300 kRad/yr at the first vertex detector layer.

\begin{center}
\includegraphics[width=8cm]{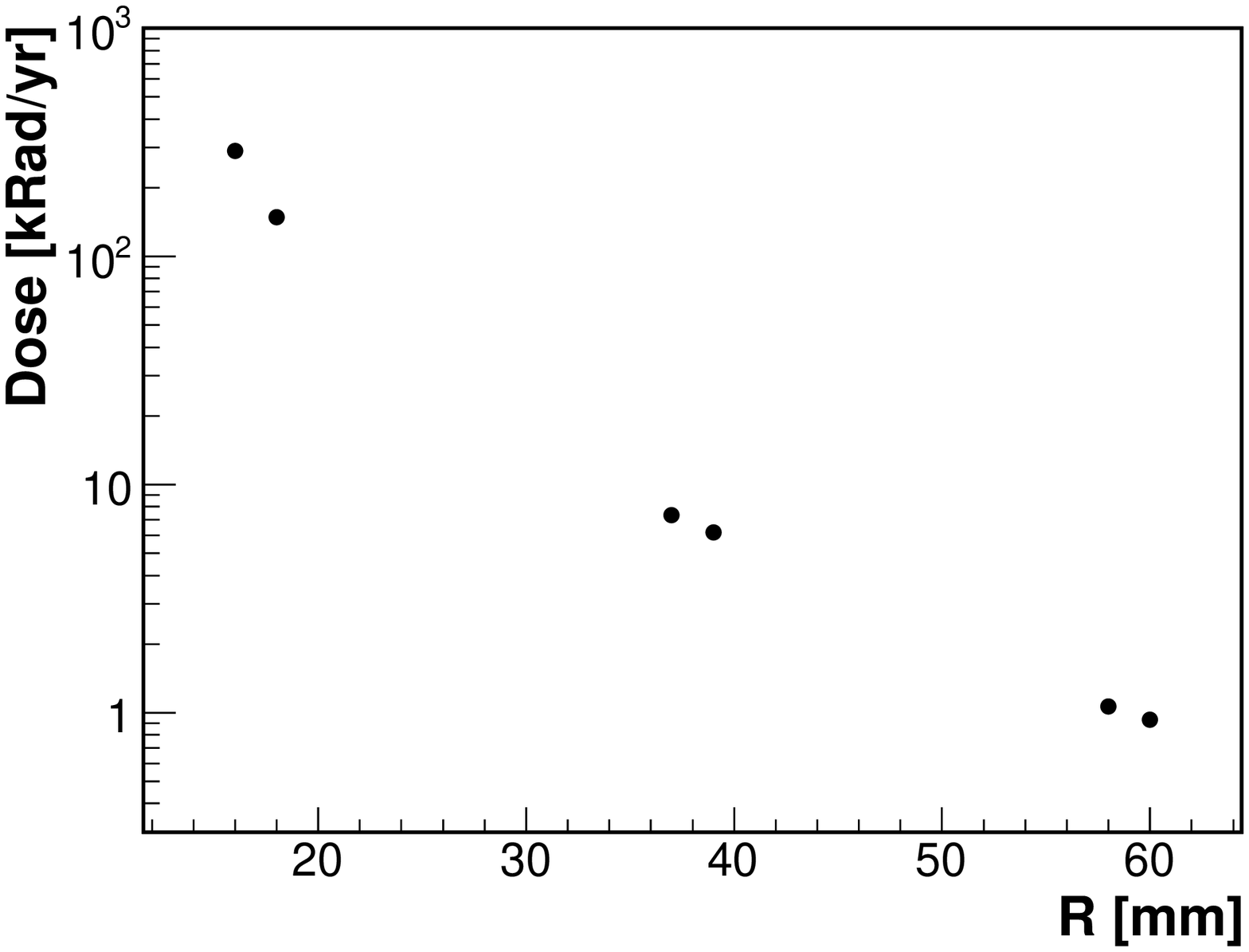}
\figcaption{\label{fig:CEPCVXDTID} TID at the vertex detector layers.}
\end{center}

\section{Conclusion}

The backgrounds induced by the beamstrahlung have been studied with both analytical and simulation methods. The hit density on the first vertex detector layer is about 0.2 hits/cm$^2$/BX and the detector occupancy is well below 0.5\%. The annual values for NIEL and TID, representing the levels of the radiation damage in silicon detector, are estimated to be $10^{11}~\mathrm{1~MeV}~ n_{eq}/\mathrm{cm}^2$ and $\sim$300 kRad, respectively. Further studies on the beamstrahlung effects and other sources of detector backgrounds, as well as detailed evaluation of their impacts on physics measurements will be pursued in future.
\\

\acknowledgments{The authors would like to thank WANG Dou, GENG Huiping, YUE Teng and WANG Yiwei for fruitful discussions on accelerator physics, and MA Mingming for helping to generate simulation samples of large statistics.}

\end{multicols}

\vspace{10mm}

\begin{multicols}{2}

\bibliography{BeamstrahlungBib}

\end{multicols}

\clearpage

%\end{CJK*}
\end{document}